\documentclass[english,aps,prl,nofootinbib]{revtex4-2}
\usepackage{lmodern}
\usepackage[LGR,T1]{fontenc}
\usepackage[latin9]{inputenc}
\usepackage{amsmath}
\usepackage{amssymb}
\usepackage{graphicx}
\usepackage{color}
\usepackage{soul}

\makeatletter

\DeclareRobustCommand{\greektext}{%
  \fontencoding{LGR}\selectfont\def\encodingdefault{LGR}}
\DeclareRobustCommand{\textgreek}[1]{\leavevmode{\greektext #1}}
\ProvideTextCommand{\~}{LGR}[1]{\char126#1}

\usepackage{enumerate}

\makeatother

\usepackage{babel}
\begin{document}
\title{The complete exterior spacetime of spherical Brans-Dicke stars}
\author{Bertrand Chauvineau$\,$}
\email[\ ]{bertrand.chauvineau@oca.eu}

\affiliation{Universit\'e C\^ote d'Azur, Observatoire de la C\^ote d\textquoteright Azur,
CNRS, Laboratoire Lagrange, Nice cedex 4, France}
\author{Hoang Ky Nguyen$\,$}
\email[\ ]{hoang.nguyen@ubbcluj.ro}

\affiliation{Department of Physics, Babe\c{s}-Bolyai University, Cluj-Napoca 400084,
Romania}
\date{\today}
\begin{abstract}
We derive the complete expression for the Brans Class I exterior spacetime
explicitly in terms of the energy and pressures profiles of a stationary
spherisymmetric gravity source. This novel and generic expression
is achieved in a \emph{parsimonious} manner, requiring only a subset
of the Brans-Dicke field equation and the scalar equation. For distant
orbiting test particles, this expression promptly provides a simple,
closed and exact formula of the \textgreek{g} Eddington parameter,
which reads $\gamma_{\,\text{exact}}=\frac{\omega+1+\left(\omega+2\right)\Theta}{\omega+2+\left(\omega+1\right)\Theta}$,
where $\Theta$ is the ratio of the star's ``total pressure'' integral
over its energy integral. This \emph{non-perturbative} result reproduces
the usual Post-Newtonian $\frac{\omega+1}{\omega+2}$ expression in
the case of a ``Newtonian star'', in which the pressure is negligible
with respect to the energy density. Furthermore, it converges to the
General Relativity value ($\gamma_{\,\text{GR}}=1$) as the star's
equation of state approaches that of ultra-relativistic matter (in
which case $\Theta$ approaches 1), a behavior consistent with broader
studies on scalar-tensor gravity. Our derivation underscores the essence
of these results involving (1) the key relevant portion of the Brans-Dicke
field equations, (2) the uniqueness of the Brans Class I vacuum solution
for the non-phantom action, viz. $\omega>-3/2$, and (3) the involvement of only two free parameters in this solution, hence requiring two quantities (energy and pressure integrals) of the mass source to fully characterize the solution. From a practical standpoint, it elucidates how a given stellar interior structure model determines
the star's exterior gravitational field and impacts the motions of
light objects (such as planets and accretion disks) orbiting it.
\end{abstract}
\maketitle
General Relativity (GR) proved to be a very powerful gravity theory,
able to account for many physical and astrophysical issues. This includes Solar System dynamics \citep{Will-livingRev,Will-book}, binary pulsars
dynamics \citep{Will-livingRev,HulseTaylor}, daily Global Positioning
System (GPS) technological application \citep{Ashby,Bertolami}, dynamics
\citep{Ghez,Riess,Perlmutter,Genzel} and optical effects (shadow)
\citep{Akiyama-2019,Akiyama-2021,Akiyama-2022} in supermassive black
holes' neighborhood, and the ability of GR to account for many high
energy astrophysics phenomena, like quasar or active galactic nuclei
physics \citep{Gabian} and binary black holes merging \citep{Abbott}.
Despite these many successes, the quest to quantize gravity and unify
it with other known interactions has spurred significant efforts,
which produce, such as via superstring theories, (classical) gravity
theory in the low energy regime that qualitatively depart from GR
by the presence of a scalar partner to the metric field. More specifically,
the gravitational sector often achieves a Brans-Dicke (BD) structure
\citep{BransDicke,Dicke-1962,Brans-1962}, the BD theory being a special
case of scalar-tensor (ST) theories \citep{Bergmann,Nordtvedt,Wagoner}.
In brief, while the metric of an ST theory describes the spacetime
geometry in a GR similar way, the scalar field determines the local
value of Newton's gravitational ``constant'', which then depends on
the spatial location and varies with time. These theoretical circumstances
have stimulated a significant revival of interest in ST theories in
the past three decades, despite the aforementioned successes of GR.
(See for instance \citep{FujiiMaeda,Faraoni,Capozziello} for overviews
on BD/ST theories and their links with attempts to quantify gravity.)
Hence, we revisit some aspects of BD gravity in this letter.\vskip4pt

In far field regions, the ST nature of gravity leaves its imprint
on the Eddington parameters $\left(\beta,\,\gamma\right)$, which
determine the first Post-Newtonian (PN) terms of the metric in the
PN approximation. On the other hand, the description of the strong
field region is generally by far more complex. The Eddington parameters
are defined in such a way that their GR values are $\left(\beta_{\,\text{GR}},\,\gamma_{\,\text{GR}}\right)=\left(1,1\right)$.
Therefore, the closer to unity are $\left(\beta,\,\gamma\right)$
for a specific theory, the closer is the (far) metric from the GR
one, and are the predicted particle's orbit to GR's predictions. Each
of the ST theories previously specified is characterized by a specific
$\omega\left(\Phi\right)$ function. The ST Eddington parameters reads
\citep{Will-livingRev}
\begin{equation}
\left\{ \begin{array}{l}
\beta=1+\frac{\Phi_{0}\omega_{0}^{\prime}}{\left(2\omega_{0}+3\right)\left(2\omega_{0}+4\right)^{2}}\\
\gamma=\frac{\omega_{0}+1}{\omega_{0}+2}
\end{array}\right.\label{usual Edd's}
\end{equation}
where $\omega_{0}=\omega\left(\Phi_{0}\right)$ and $\omega_{0}^{\prime}=\omega^{\prime}\left(\Phi_{0}\right)$,
$\Phi_{0}$\ being the value of the scalar at the considered event.
The experimental tests constrain $\left(1-\gamma\right)$ at the $10^{-5}$
level (see chapter 7 of Ref. \citep{Will-book}), and require acceptable
ST theories to satisfy $\omega_{0}>4\times10^{-5}$. (Let us remind
that the $\omega_{0}<-3/2$ values correspond to a ``phantom'' kinetic
energy for the scalar field, a case that could be deemed unphysical.
Accordingly, the $\omega_{0}\ll-1$ possibility is usually discarded.)
If gravity would be of ST nature, it could seem unnatural that it
is adjusted in such a way that it mimics so closely GR. However, it
has been shown in the nineties that a significant subset of ST theories
obey an ``attractor mechanism'' in which the Universe's expansion
drives the scalar field to a value for which $\omega\left(\Phi\right)$
diverges, which finally imposes the theory to mimic GR \citep{Damour-1993a,Damour-1993b}.\vskip4pt

The BD theory \citep{BransDicke} corresponds to the case where $\omega$
is $\Phi$ independent, i.e. is constant. The theory offers the advantage
that the exact and general vacuum static spherical solution can be
written in explicit form. It is the Brans Class I solution in the
$\omega>-3/2$ case. This solution, which depends on two parameters,
is the BD counterpart of the GR Schwarzschild solution, which depends
on just one parameter. However, there is a strong difference between
BD and GR: the Birkhoff theorem no longer applies in BD gravity. Therefore,
the Brans Class I solution only describes the external gravitational
field of \textit{static} spherical stars.\vskip4pt

Specifying to the case where Brans Class I is sourced by a Newtonian star (a weak field star in which the pressure $p$ is negligible
with respect to the energy density $\epsilon$)
and Taylor expanding the exterior metric, Brans and Dicke
essentially found \footnote{One should note that Brans and Dicke did not explicitly provide the
Eddington parameters in their 1961's paper, but the prerequisites
were available therein through Eqs. (29), (32) and (34) in Ref. \citep{BransDicke}.}
\begin{equation}
\left\{ \begin{array}{l}
\beta_{\,\text{BD}}=1\\
\gamma_{\,\text{BD}}=\frac{\omega+1}{\omega+2}
\end{array}\right.\label{BD's Edd's}
\end{equation}
in their seminal paper \citep{BransDicke}, in accordance with \eqref{usual Edd's}.
However, considering asymptotically flat solutions, the Taylor expansion
of the metric, in its form involving the Eddington parameters, is
generic in far regions of the spacetime. It is then also relevant
in the case of a compact and static spherical star. The purpose of
this letter is to provide the exact and readily applicable expressions
of the Brans Class I solution parameters, and of the induced expression
of $\gamma$, in terms of the stellar internal structure.\vskip4pt

The (Jordan-frame) BD field equations read
\begin{align}
R_{ab}-\frac{\omega}{\Phi^{2}}\partial_{a}\Phi\partial_{b}\Phi-\frac{1}{\Phi}\partial_{a}\partial_{b}\Phi+\Gamma_{ab}^{c}\partial_{c}\ln\Phi & =\frac{8\pi}{\Phi}\left(T_{ab}-\frac{\omega+1}{2\omega+3}Tg_{ab}\right)\label{BD (a,b) field eq}\\
\partial_{a}\left(\sqrt{-g}g^{ab}\partial_{b}\Phi\right) & =\frac{8\pi}{2\omega+3}T\sqrt{-g}.\label{BD scalar field eq}
\end{align}
Let us consider a static spherisymmetric spacetime in isotropic coordinates
\begin{equation}
ds^{2}=-A\left(r\right)dt^{2}+B\left(r\right)\left[dr^{2}+r^{2}\left(d\theta^{2}+\sin^{2}\theta d\varphi^{2}\right)\right]\label{spherical metric}
\end{equation}
with a scalar $\Phi\left(r\right)$. The stationarity and spherisymmetry
also imposes the stress tensor to have the form (let us spot that
stationarity prevents the existence of radial heat fluxes)
\begin{align}
\left(T_{a}^{b}\right) & =\text{diag}\bigl(-\epsilon,p_{\Vert},p_{\bot},p_{\bot}\bigr)\label{stress tensor}
\end{align}
where $\bigl(\,\epsilon,p_{\Vert},p_{\bot}\bigr)$ are three $r$--dependent
functions. Thence, the most general gravity source differs from the
perfect fluid by the only anisotropy between the radial and orthogonal
pressures $p_{\Vert}$ and $p_{\bot}$. The \eqref{BD (a,b) field eq}--\eqref{BD scalar field eq}
system yields four independent non-trivial equations. These four equations
are needed for a complete stellar internal structure description,
but one defers this for dedicated studies. For what is targeted here,
it turns out that
only the scalar equation and the $\left(00\right)$ component of Eq.
\eqref{BD (a,b) field eq} are required. These two equations read
respectively
\begin{align}
\left(r^{2}\sqrt{AB}\Phi^{\prime}\right)^{\prime} & =\frac{8\pi}{2\omega+3}\Bigl[-\epsilon+p_{\Vert}+2p_{\bot}\Bigr]r^{2}\sqrt{AB^{3}}\label{BD eq(Phi)}\\
\left(r^{2}\Phi\sqrt{\frac{B}{A}}A^{\prime}\right)^{\prime} & =16\pi\Bigl[\,\epsilon+\frac{\omega+1}{2\omega+3}\bigl(-\epsilon+p_{\Vert}+2p_{\bot}\bigr)\Bigr]r^{2}\sqrt{AB^{3}}\label{BD eq(A)}
\end{align}
where relativistic units are used. These equations offer the advantage
of having both their left hand sides in exact derivative forms. \emph{Outside
the star}, the scalar-metric is the Brans Class I solution (which
of course satisfies the full \eqref{BD (a,b) field eq}--\eqref{BD scalar field eq}
system), which reads \citep{Brans-1962}
\begin{equation}
\left\{ \begin{array}{l}
A=\left(\frac{r-k}{r+k}\right)^{\frac{2}{\lambda}}\\
B=\left(1+\frac{k}{r}\right)^{4}\left(\frac{r-k}{r+k}\right)^{2-2\frac{\Lambda+1}{\lambda}}\\
\Phi=\left(\frac{r-k}{r+k}\right)^{\frac{\Lambda}{\lambda}}
\end{array}\right.\ \ \ \ \ \text{for }r\geqslant r_{*}\label{BC1}
\end{equation}
where $r_{\ast}$ is the star's radius, and 
\begin{equation}
\lambda^{2}=\left(\Lambda+1\right)^{2}-\Lambda\left(1-\frac{\Lambda}{2}\omega\right)\label{relation}
\end{equation}
Since $\lambda$ and $\Lambda$ are linked by \eqref{relation}, this
solution involves two independent parameters only, which one chooses to
be $\left(k,\,\Lambda\right)$. In remote spatial regions, the Taylor
expansion of the metric yields the gravitational mass and the Eddington
parameters. They read
\begin{eqnarray}
m & = & \frac{2k}{\lambda}\label{BD's ADM mass}\\
\beta_{\,\text{exact}} & = & 1\label{BD's beta}\\
\gamma_{\,\text{exact}} & = & 1+\Lambda\label{BD's gamma}
\end{eqnarray}
where $1+\Lambda$ is a priori \textit{not fixed} to be $\frac{\omega+1}{\omega+2}$
(or anything else), i.e. $\Lambda$ not fixed to be $\frac{-1}{\omega+2}$,
which was implicitly recognized by Brans and Dicke who explicitly
wrote $\Lambda\simeq-1/(\omega+2)$ in their seminal paper (see Eq.
(34) of Ref. \citep{BransDicke}). The exact link between $\Lambda$ and $\omega$ can be obtained by integrating Eqs. \eqref{BD eq(Phi)} and \eqref{BD eq(A)} from the star's center up to a distance $r$ that resides outside of the star, i.e. $r>r_\ast$, which is the approach to be adopted henceforth. The fact that Brans Class I involves two independent parameters explains why just two field equations are needed for achieving the link. \footnote{It is worth mentioning that the two equations which yield, in the
relevant approximation scheme, the Poisson equation (and the expression
of the effective gravitational constant in terms of the scalar field
and $\omega$) also fully determine the exact form of the external
gravitational field in the spherical case, once the pressures and
energy profiles are known. (Determining these profiles would require
the full BD equations system, completed by relevant equations of state.)}\vskip4pt

Let us integrate Eqs. \eqref{BD eq(Phi)} and \eqref{BD eq(A)} from
the star's center to a distance $r>r_{\ast}$. The $\left(A,B,\Phi\right)$
functions are then given by \eqref{BC1} at $r$. For $r>r_{*}$,
both $r^{2}\sqrt{AB}\Phi^{\prime}$ and $r^{2}\Phi\sqrt{\frac{B}{A}}A^{\prime}$
terms that enter the left hand sides of \eqref{BD eq(Phi)} and \eqref{BD eq(A)}
are $r-$independent, since the right hand sides of these equations
vanish in the \emph{exterior} vacuum. On the other hand, regularity
conditions inside the star impose $\Phi^{\prime}\left(0\right)=A^{\prime}\left(0\right)=0$
(for having no conic singularity) and finite values of the fields
themselves. The calculation yields
\begin{align}
\frac{k\,\Lambda}{\lambda} & =\frac{4\pi}{2\omega+3}\int_{0}^{r_{\ast}}dr\,r^{2}\sqrt{AB^{3}}\Bigl[-\epsilon+p_{\Vert}+2p_{\bot}\Bigr]\label{BD eq(Phi) integr}
\end{align}
and
\begin{align}
\frac{k}{\lambda} & =\frac{4\pi}{2\omega+3}\int_{0}^{r_{\ast}}dr\,r^{2}\sqrt{AB^{3}}\Bigl[(\omega+2)\epsilon+(\omega+1)\bigl(p_{\Vert}+2p_{\bot}\bigr)\Bigr].\label{BD eq(A) integr}
\end{align}
Let us note that $r^{2}\sqrt{AB^{3}}$ is the square root of the determinant
of the metric, up to the $\sin\theta$ term. (Accordingly, the integrals
in the right hand sides of Eqs. \eqref{BD eq(Phi) integr} and \eqref{BD eq(A) integr}
are invariant through radial coordinate transformations, if $r^{2}\sqrt{AB^{3}}\sin\theta$
is replaced by $\sqrt{-g}$.) One defines the energy's and pressures'
integrals by
\begin{eqnarray}
E^{\ast} & = & 4\pi\int_{0}^{r_{\ast}}dr\,r^{2}\sqrt{AB^{3}}\,\epsilon\label{star's energy integr}\\
P_{\Vert}^{\ast} & = & 4\pi\int_{0}^{r_{\ast}}dr\,r^{2}\sqrt{AB^{3}}\,p_{\parallel}\label{star's rad pressure integr}\\
P_{\bot}^{\ast} & = & 4\pi\int_{0}^{r_{\ast}}dr\,r^{2}\sqrt{AB^{3}}\,p_{\perp}.\label{star's orthog pressure integr}
\end{eqnarray}
Inserting in \eqref{BD eq(Phi) integr} and \eqref{BD eq(A) integr},
we obtain
\begin{equation}
\frac{k}{\lambda}= \frac{\omega+2}{2\omega+3}\,E^{*}+\frac{\omega+1}{2\omega+3}\bigl(P_{\parallel}^{*}+2P_{\bot}^{*}\bigr)\label{k/lambda}
\end{equation}
and
\begin{equation}
\Lambda=\frac{-E^{*}+P_{\parallel}^{*}+2P_{\perp}^{*}}{(\omega+2)E^{*}+(\omega+1)\bigl(P_{\parallel}^{*}+2P_{\bot}^{*}\bigr)}\label{Lambda}
\end{equation}
which, together with \eqref{BC1} and \eqref{relation}, provide a
complete expression for the exterior spacetime and scalar field of
a spherical BD star. To the best of our knowledge, this prescription
has not been explicitly documented in the literature.\vskip4pt

For a perfect fluid, $p_{\Vert}=p_{\bot}\equiv p$, thence $P_{\Vert}^{\ast}=P_{\bot}^{\ast}\equiv P$.
The equations \eqref{relation}, \eqref{BD eq(Phi) integr} and \eqref{BD eq(A) integr}
fully determine the exterior solution \eqref{BC1} once the integrals
\eqref{star's energy integr}--\eqref{star's orthog pressure integr}
are known, with these integrals being fixed by the stellar internal
structure model. This explicitly determines the particles' motion
outside the star, in both the remote and close to the star regions.
Let us spot that, from \eqref{BD's ADM mass}, the integral \eqref{BD eq(A) integr}
is exactly half of the mass of the star.\vskip4pt

For dynamics in the remote region, one further finds, combining Eqs. \eqref{BD's gamma} and \eqref{Lambda}
\begin{equation}
\gamma_{\,\text{exact}}=\frac{\omega+1+\left(\omega+2\right)\Theta}{\omega+2+\left(\omega+1\right)\Theta}\label{exact gamma}
\end{equation}
where $\Theta\equiv\left(P_{\Vert}^{\ast}+2P_{\bot}^{\ast}\right)/E^{\ast}$
is the ratio of the ``total pressure integral'' $P_{\Vert}^{\ast}+2P_{\bot}^{\ast}$
to the energy integral. To our knowledge, the closed-form expression
\eqref{exact gamma} is presented here for the first time.\vskip4pt

\begin{figure}[!t]
\noindent \begin{centering}
\includegraphics[scale=0.58]{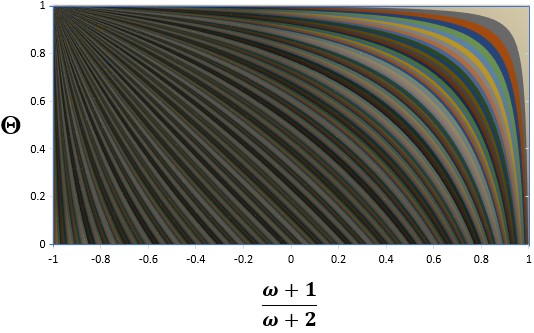}
\par\end{centering}
\caption{\label{fig:gamma-countour}Contour plot of $\gamma_{\text{exact}}$
in terms of $\varTheta$ and $\frac{\omega+1}{\omega+2}$, for the
physical range of $\varTheta\in[0,1]$ and $\omega\in(-3/2,+\infty)$.
The latter is translated to $\frac{\omega+1}{\omega+2}\in(-1,1)$.
Contours are equally spaced in 0.005 increments. Note that for a given
contour, the corresponding value of $\gamma_{\,\text{exact}}$ equals
to that on the abscissa axis where the contour intersects it. A value
of $\gamma_{\,\text{exact}}$ approximately 1 can infer that $\frac{\omega+1}{\omega+2}\approx1$
(i.e., $\omega\gg1$) or $\varTheta\approx1$ (i.e., ultra-relativistic
matter). }
\end{figure}

In the case of a perfect fluid Newtonian star, $p\ll\epsilon$, which
yields $\Theta\ll1$. Therefore, the usual $\gamma_{\,\text{BD}}$
expression in \eqref{BD's Edd's} is recovered. On the other hand,
if the pressures cannot be neglected, $\gamma_{\,\text{exact}}$ carries
mixed information on the gravity theory (through $\omega$ in this
BD example) and on the star's physical properties. This can be seen
in the contour plot in Figure \ref{fig:gamma-countour} which shows
the value of $\gamma_{\,\text{exact}}$ in terms of $\Theta$ and
$\frac{\omega+1}{\omega+2}$. For example, with $\omega=2$ and $\Theta=0.8$,
one would obtain $\gamma_{\,\text{exact}}=0.96875$, while the weak
field expression yields 0.75.\vskip4pt

Let us stress that since this is also true in the perfect fluid case,
the claim has nothing to do with a possible anisotropic pressure.
Despite the fact that a linear barotropic equation of state (EoS)
$p\propto\epsilon$ (considering the perfect fluid case) describes
a medium of infinite extension (no surface for a linear barotropic
medium), one can contemplate the case of a star the matter of which
is close to the ultra-relativistic EoS, i.e. $p$ is close to $\epsilon/3$
through the whole star apart from a layer close to its surface. In
this case, $\Theta$ is close to $1$, in such a way that $\gamma_{\,\text{exact}}$
is close to $1$, i.e. to the GR value. Let us stress that this point
does not dependent on the $\omega$ value. To some extent, the BD theory mimics GR despite $\omega$ is finite, because of the internal structure of the star.\vskip4pt

It should be noted that the previous conclusions may have been deducible from Ref.~\citep{DEF-1992}, which considered the general multi-tensor-scalar case in great details. Nevertheless, we have provided here, \emph{for the first time}, the explicit and ready-for-use compact formula \eqref{exact gamma} for the $\gamma$ parameter in the BD case. Moreover, our derivation of \eqref{exact gamma} was direct, transparent, and parsimonious, underscoring the essential ingredients involved. Let us also point out that, since the $(k,\,\Lambda)$ constants are exactly known, not only $\gamma$ but also {\it any} PN parameter can be determined by expanding the metric functions $A$ and $B$ in Eq.~\eqref{BC1} accordingly. In addition, Eqs.~\eqref{k/lambda} and \eqref{Lambda} (with the aid of \eqref{relation}) also elucidate how a given stellar interior model directly impacts the orbital dynamics of bodies orbiting at \emph{intermediate} distances, where the use of the PPN formalism is questionable since the gravitational field is not weak at these distances. Indeed, the \eqref{star's energy integr}--\eqref{star's orthog pressure integr} integrals also fully and exactly determine the parameters of the Brans Class I solution \eqref{BC1}, which then also allow to determine orbital dynamics up to the strong field region of the mass source.\vskip4pt

The statement above can be quantified to dictate the extent for a given value of $\gamma$ to fix the BD theory, i.e. constraining the $\omega$ value.  While the usual weak field expression \eqref{BD's Edd's} completely fixes $\omega$ as soon as $\gamma$ is known, \eqref{exact gamma} yields instead
\begin{equation}
\omega =\frac{1}{\delta}\cdot\frac{1-\Theta}{1+\Theta}
-\frac{2+\Theta}{1+\Theta} \label{omega-vs-Theta}
\end{equation}
where one has defined $\delta=1-\gamma_{\,\text{exact}}$. Thence, because of the $\Theta$ dependence, the measurement of $\gamma$ (viz. $\delta$) alone is not sufficient to fix the underlying BD theory, even if $(\gamma-1)$ is close to zero.\vskip4pt
Specifying the small $\Theta$ case, first order expanding Eq. \eqref{omega-vs-Theta} with respect to $\Theta$ yields
\begin{equation}\omega =\frac{1}{\delta}\left(1-2\Theta\right)-2+\Theta+\dots
\label{omega Theta10}
\end{equation}
which, when $\delta\approx0$, is further simplified to $\omega \approx \frac{1}{\delta}-\frac{2\Theta}{\delta}$. The last expression shows that for small $\delta$, a weak $\Theta$ value would not affect the main $1/\delta$ estimate for $\omega$. Specifically for the Solar System, the numerical value of $\Theta$ can be calculated using Christensen-Dalsgaard et al's Solar internal structure model \citep{C-D Sun's model}. One obtains $\Theta_\text{\,Sun}=3.43\times10^{-6}$ \citep{Strugarek-perso}, which confirms that $\Theta$ has no material impact on determining the BD parameter by Solar System experiments.\vskip4pt

On the other hand, considering neutron star systems where $\Theta$ is known to be significant, the contribution $2\Theta/\delta$ would be substantial as compared with the $1/\delta$ estimate of $\omega$ when $\Theta=0$. The explicit $\Theta$ dependence of $\gamma$, and/or the pair $(\Lambda,k)$, would then become of primary interest when the dynamics of such systems is used for constraining gravity theories.\vskip8pt

As a final remark, let us stress that the calculation presented here
reveals an important consequence. It shows that, even in the spherical
case, motions around a compact star do not only carry information
on the gravity theory, but that this information is mixed with features
of the star's structure beyond its only mass. The point has some acquaintance
with the loss of the Birkhoff theorem in BD. Indeed, this loss means
that the pulsations (or the collapse) of a non static spherical star
affect orbital motions around it, which therefore are no longer just
determined by the star's mass (for a given $\omega$ parameter, i.e.
BD theory). The notable point here is that, also in the static case,
the sole mass knowledge does not permit to know what PN orbital motions
around the star are. Knowing more is needed.\vskip8pt

\emph{Acknowledgments}---The authors thank the anonymous referee for their constructive comments. BC thanks Antoine Strugarek and HKN thanks Valerio Faraoni for their helpful correspondences.

\end{document}